\documentclass[a4paper,twocolumn]{revtex4}
\def\figps#1{\vbox to 6.5cm{
\includegraphics{#1}
\vfill}}
\begin{document}
\title{Sampling along reaction coordinates with the Wang-Landau method}
\author{F. Calvo}
\affiliation{Laboratoire de Physique Quantique, IRSAMC,
Universit\'e Paul Sabatier,
118 Route de Narbonne, F31062 Toulouse Cedex, France}
\begin{abstract}
The multiple range random walk algorithm recently proposed by Wang and Landau
[2001, Phys. Rev. Lett., 86, 2050-2053] is adapted to
the computation of free energy profiles for molecular systems along reaction
coordinates. More generally, we show how to extract partial averages in various
statistical ensembles without invoking simulations with constraints, biasing
potentials or unknown parameters. The method is illustrated on a model
10-dimensional potential energy surface, for which analytical results are
obtained. It is then applied to the potential of mean force associated with the
dihedral angle of the butane molecule in gas phase and in carbon tetrachloride
solvent. Finally, isomerisation in a small rocksalt cluster, (NaF)$_4$, is
investigated in the microcanonical ensemble, and the results are compared to
those of parallel tempering Monte Carlo.
\end{abstract}
\maketitle

\section{Introduction}
\label{sec:intro}

The calculation of free energies, or free energy differences, is a difficult
task for simulation. Free energies cannot be conveniently expressed as
averages, and because of this specific answers have been proposed, including
the well known thermodynamic perturbation and integration methods
\cite{allen,frenkel}, but also biasing potentials \cite{vandevondele} and
umbrella sampling \cite{torrie1,torrie2}, constraint dynamics in the so-called
``blue-moon'' ensemble \cite{carter,paci}, adiabatic switching \cite{watanabe}
and reversible-scaling \cite{yip}, and the recent adiabatic molecular
dynamics method \cite{rosso1,rosso2}. Given two states A and B in
configuration space, these techniques allows one to get a thermodynamic picture
of the reaction pathway between A and B, but also to estimate transition rates.
When the reaction coordinate is not well defined, it is still possible to
calculate such rates using the transition path sampling approach
\cite{tsp,tspwales}.

Most of the above methods suffer from important limitations when it comes to a
practical situation. One must either find by trial and error a suitable guiding
potential (the multicanonical ensemble sampling \cite{multicano} can help in
this search), or deal with the specific transformation rules required by
constraint dynamics, which can often be cumbersome in the case of angle
coordinates. The adiabatic free energy dynamics method \cite{rosso1,rosso2}
also makes use of variable transformation and particular integrators.

Recently, Wang and Landau have proposed a very simple algorithm to compute the
density of states of spin systems \cite{wl}. Their multiple range random walk
method has since been extended to spin glasses \cite{wl2}, continuous fluids
\cite{depablo}, proteins \cite{depablo2} and polymer fluids \cite{depablo3}. A
combination with the kinetic Monte Carlo algorithm has also been made by
Schulz, Binder, and M\" uller \cite{schulz}. Although this method was
originally conceived to calculate a microcanonical quantity, constant
temperature properties can be recovered using appropriate Laplace
transformations. The main interest of the algorithm is that it does not require
any information {\em a priori}, other than a suitable choice of reaction
coordinate. Instead it converges to the expected
solution with a greater accuracy at each new iteration. The goal of this
paper is to propose a simple way to use the Wang-Landau method in the context
of free energy profiles. As will be shown below, the limitations mentioned
earlier essentially vanish, making the method really useful in various
statistical ensembles.

The article is organised as follows. In the next section, we give the main
ideas of the Wang-Landau algorithm in the framework of free energy
calculations. In section~\ref{sec:res}, we illustrate this method on three
examples. Firstly, a
simple analytical 10-dimensional potential energy surface is investigated and
the simulation results are compared with analytical data. Secondly, the free
energy profile of the butane molecule along the dihedral angle coordinate is
calculated in gas phase and CCl$_4$ solvent, and compared with previous works
on the same system \cite{depaepe}. Thirdly, we study the cube-ring
isomerisation in a small rocksalt cluster, namely (NaF)$_4$, in the
microcanonical ensemble. We briefly summarise and conclude in
section~\ref{sec:ccl}.

\section{The Wang-Landau method for free energy profiles}
\label{sec:met}

We consider a classical N-dimensional system in a given statistical ensemble
characterized by a equilibrium distribution $\rho$. This distribution is a
function of the cartesian coordinates ${\bf R} = \{ x_i, y_i, z_i \}$, and
also of external parameters such as temperature or total energy. A reaction
coordinate $\lambda({\bf R})$ is constructed, and we look for the probability
density $p(\lambda_0)$ for the reaction coordinate to take the particular
value $\lambda_0$:
\begin{equation}
p(\lambda_0) = \langle \delta [ \lambda({\bf R}) - \lambda_0 ] \rangle.
\label{eq:pl0}
\end{equation}
In this equation, $\langle A\rangle$ denotes the average value of $A$ in the
statistical ensemble characterised by $\rho$:
\begin{equation}
\langle A\rangle = \left. \int A({\bf R}) \rho({\bf R})d{\bf R} \right/
\int \rho({\bf R})d{\bf R}.
\label{eq:ava}
\end{equation}
The original multiple range random walk method of Wang and Landau \cite{wl}
consists of performing a Monte Carlo simulation using a Metropolis acceptance
probability with a dynamical weight $g(E)$, where $E({\bf R})$ is the energy of
configuration ${\bf R}$:
\begin{equation}
\mbox{acc} ({\bf R}_{\rm old} \to {\bf R}_{\rm new}) = \min \left[ 1,
\frac{g(E_{\rm old})}{g(E_{\rm new})} \right].
\label{eq:wle}
\end{equation}
For each visited state with energy $E$, $g$ is scaled by a constant factor $f$:
$g(E) \to f\times g(E)$. The scaling factor $f$ is initially quite large
$(\sim 2$--$2.5)$, and is periodically reduced toward one. $g$ is initially set
to one for all values of $E$. As shown by Wang and Landau \cite{wl}, this
algorithm smoothly converges to a flat probability distribution when $f\to 1$,
meaning that the function $g$ converges to the density of states $\Omega$ of
the system.

Since $\Omega(E)$ and $p(\lambda)$ play similar roles, the same
ideas can be applied to the problem of free energy profiles along reaction
coordinates. Let $g(\lambda)$ be a function initially set to one in the whole
range of accessible values of $\lambda$. For practical and numerical purposes,
it turns out that it is more convenient to work with $s(\lambda) = \ln g
(\lambda)$ instead of $g$. We perform a Monte Carlo simulation using the
following Metropolis acceptance rule:
\begin{eqnarray}
\mbox{acc}({\bf R}_{\rm old}\to {\bf R}_{\rm new}) &=&\displaystyle
\min\left[ 1, \frac{\rho({\bf R}_{\rm new})}{\rho({\bf R}_{\rm old})} \times
\frac{g(\lambda_{\rm old})}{g(\lambda_{\rm new})} \right]\nonumber \\
&=& \displaystyle \min\left[ 1,
\frac{\rho({\bf R}_{\rm new})}{\rho({\bf R}_{\rm old})} \times
\frac{e^{-s(\lambda_{\rm new})}}{e^{-s(\lambda_{\rm old})}} \right],
\label{eq:ert}
\end{eqnarray}
with $\lambda_{\rm new} = \lambda({\bf R}_{\rm new})$ and $\lambda_{\rm old} =
\lambda({\bf R}_{\rm old})$. After the new configuration is visited, the
corresponding $s$ is updated: $s(\lambda) \to s(\lambda) + \alpha$. The
parameter $\alpha = \ln f$ is progressively decreased to zero at each new
iteration. The histogram $h(\lambda)$ of visited values of $\lambda$ is
considered flat when $h(\lambda) > \varepsilon \langle h(\lambda)\rangle$ for
all $\lambda$. The parameter $\varepsilon$ is usually taken between 0.5 and
0.99. After a large number of MC steps and some iterations, $g=e^s$ gets
close to the probability density $p$ defined in equation (\ref{eq:pl0}).
As noted by Wand and Landau \cite{wl}, the continuous change in the
statistical weight in Eqn.~(\ref{eq:wle},\ref{eq:ert}) strictly violates
detailed balance. Actually, detailed balance is satisfied only after the
function $s(\lambda)$ becomes nearly constant, and within an accuracy of
order of magnitude $\alpha$ \cite{wl}. In fact, as is well known in
statistical mechanics, detailed balance in only a {\em sufficient}\/
condition for the Monte Carlo chain to be Markovian \cite{allen,frenkel}.

In the canonical ensemble at temperature $T=1/k_B \beta$, the equilibrium
distribution is given by the Boltzmann factor $\rho({\bf R}) = \exp[ - \beta
V({\bf R})]$, where $V$ is the potential energy of configuration ${\bf R}$.
Introducing $\Gamma(\lambda) = - s(\lambda) / \beta$, equation (\ref{eq:ert})
can be rewritten as
\begin{equation}
\mbox{acc}({\bf R}_{\rm old}\to {\bf R}_{\rm new}) = \min[1, \exp(-\beta
\Delta F)],
\label{eq:bdf}
\end{equation}
with $\Delta F = F({\bf R}_{\rm new}) - F({\bf R}_{\rm old})$, and the Landau
free energy $F({\bf R}) = V({\bf R}) - \Gamma[\lambda({\bf R})]$. The function
$\Gamma$ converges to the potential of mean force (PMF) W:
\begin{equation}
W(\lambda) = - \frac{1}{\beta} \ln p(\lambda).
\label{eq:pmf}
\end{equation}
In the microcanonical ensemble at total energy $E$, the equilibrium density is
$\rho({\bf R}) = [E - V({\bf R})]^{N/2-1}$, and $s(\lambda)$ plays the role of
a Landau entropy. Extra variables can be included, as in the
isothermal-isobaric NPT ensemble, or the grand-canonical $\mu$VT ensemble.
Expressions similar to equation (\ref{eq:bdf}) are then found for the
acceptance probability, based on their usual forms without reaction coordinates
\cite{allen,frenkel}.

There is no other input to this algorithm than the function $g$, initially
unknown and set to one. In order to illustrate its efficiency, we have chosen
to apply the method to a variety of molecular systems.

\section{Simulation experiments}
\label{sec:res}

\subsection{Ten-dimensional model potential}

\begin{figure}[htb]
\figps{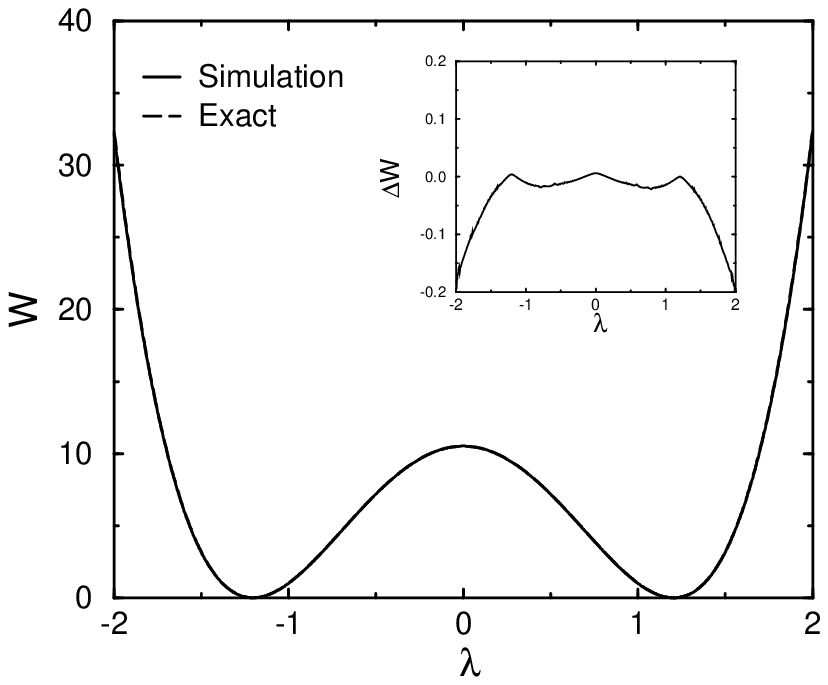}
\caption{Potential of mean force $W(\lambda)$ of the 10-dimensional potential
of equation\protect~(\ref{eq:v10}) at temperature $T=10^{-3}$, for the reaction
coordinate $\lambda=x_1$. Inset: absolute error between simulation and exact
results, $\Delta W = W_{\rm exact} - W_{\rm simul}$.}
\label{fig:pmft10}
\end{figure}

\begin{figure}[htb]
\figps{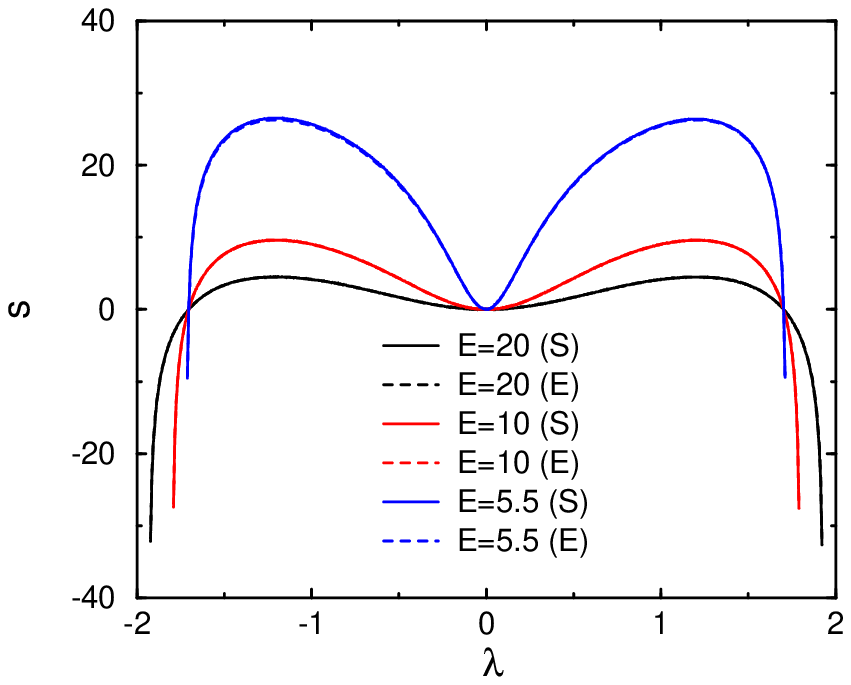}
\caption{Landau entropy $s(\lambda)$ of the 10-dimensional potential of
equation\protect~(\ref{eq:v10}) in the microcanonical ensemble at total
energies $E=20$, $E=10$, and $E=5.5$, respectively. The simulation results
are labelled (S), the exact results are labelled (E).}
\label{fig:pmfe10}
\end{figure}

Following Rosso and coworkers \cite{rosso1,rosso2}, we start to test our
method on a simple potential energy surface (PES) for which exact results can
be obtained. The expression for $V({\bf X}) = V(x_1,\dots,x_{10})$ is that of
a double-well potential
\begin{equation}
V({\bf X}) = D(x_1^2-1)^2 + \frac{1}{2} \sum_{i=2}^{10} x_i^2 + x_1
\sum_{i=2}^{10} a_i x_i.
\label{eq:v10}
\end{equation}
The parameters are $D=5$ and $a_i=1$ for all $i$. The reaction coordinate is
$\lambda=x_1$, and the PMF at temperature $T$ is found to be
\begin{equation}
W(\lambda) = D(\lambda^2-1)^2 - \frac{\lambda^2}{2} \sum_{i>1} a_i,
\label{eq:w}
\end{equation}
therefore $W$ does not depend on $T$. We have calculated the free energy
profile $W$ along the reaction coordinate $\lambda$ in the range $-2\leq\lambda
\leq 2$ using the Wang-Landau method, with 20 iterations of $1.2\times 10^6$
Monte Carlo steps, of which the first $2\times 10^5$ were removed for
equilibration. The histogram in $\lambda$ is made of 1000 bins. To illustrate
the power of the algorithm, the temperature is chosen as only $T=10^{-3}$. In
figure \ref{fig:pmft10} we have plotted the calculated PMF against the
analytical result of equation (\ref{eq:w}). The two curves are nearly
indiscernable, and inspecting the error function $\delta(\lambda) =
W_{\rm exact}(\lambda) - W_{\rm simul}(\lambda)$ in the inset of this figure
reveals how accurate the computed free energy profile is. In the microcanonical
ensemble, when the total energy $E$ is close to the saddle energy $V(0)=D$,
the true dynamics is strongly slowed down and the crossing rate between the two
wells drops to zero. The Landau entropy $s$ associated with the microcanonical
probability density can be easily calculated as
\begin{equation}
s(\lambda) = (N-3/2)\ln\left[ E - D(\lambda^2-1)^2 - \frac{\lambda^2}{2}
\sum_{i>1} a_i \right],
\label{eq:sl}
\end{equation}
up to an additive constant. The simulation was performed with the
microcanonical distribution $\rho({\bf X}) = [E-V({\bf X})]^4$, with the same
number of MC steps and iterations as above, at the total energies $E=20$, 10,
and 5.5, respectively. The results are represented in figure \ref{fig:pmfe10}.
The agreement between the numerical experiment and the exact value is again
very good, even for the most difficult case $E=5.5$. Therefore the Wang-Landau
method seems very efficient for computing free energies and related quantities
along reaction coordinates, at least in such low dimensional model problems.

\subsection{Gas phase and solvated butane}

We turn now to a more realistic system and a more complex reaction coordinate.
The {\em cis-trans}\/ isomerisation in n-butane has previously been
investigated
by several authors \cite{rosso2,depaepe,tobias}. In particular, Depaepe and
coworkers have used this molecule as a benchmark for free energy calculations
\cite{depaepe}. The accurate results obtained by these authors using umbrella
sampling and constrained dynamics in the blue-moon ensemble allow a stringent
test of the present method.

The butane molecule is modelled with a united-atom potential interacting with
the solvent molecules via a simple Lennard-Jones (LJ) potential. The total
potential energy $V$ of the system having one solute molecule and $N$ solvent
molecules is written in cartesian coordinates ${\bf R}=\{{\bf r}_1, \dots, {\bf
r}_{N+4}\}$, where $\{ {\bf r}_1, \dots, {\bf r}_4\}$, label the CH$_3$ and
CH$_2$ groups of the butane molecule, and ${\bf r}_i$, $i>4$ label the solvent
molecules. The CH$_3$ and CH$_2$ groups are connected by harmonic C-C bonds and
C-C-C bending forces. A torsion potential between the extremal methyl groups
consists of a dihedral angle term and a LJ interaction $u_b(r_{14})$. The
solvent is chosen as carbon tetrachloride, which is nonpolar and simple enough
to allow for a simple united-atom description with the LJ form. The expression
for $V$ is
\begin{equation}
V({\bf R}) = V_{\rm butane}({\bf r}_1,\dots,{\bf r}_4) + V_{\rm solvent}({\bf r
}_5, \dots, {\bf r}_{N+4}) + V_{\rm int}({\bf R}),
\label{eq:vbutane}
\end{equation}
with
\begin{equation}
V_{\rm butane}({\bf r}_1,\dots,{\bf r}_4) = \sum_{i=1}^3 \frac{k_s}{2} (
| {\bf d}_i| - d_i^*)^2 + \sum_{i=1}^2 \frac{k_b}{2}(\theta_i - \theta_i^*)^2
+ u_3 \cos (3\alpha) + u_b(r_{14}),
\label{eq:but}
\end{equation}
\begin{equation}
V_{\rm solvent}({\bf r}_5,\dots,{\bf r}_{N+4}) = \sum_{4<i<j} u_s(r_{ij}),
\label{eq:sol}
\end{equation}
\begin{equation}
V_{\rm int}({\bf R}) = \sum_{i=1}^4 \sum_{j>4} u_{bs}(r_{ij}).
\label{eq:int}
\end{equation}
In equation (\ref{eq:but}), $| {\bf d}_i|$ and $\theta_i$ are the C-C bond
lengths and C-C-C bending angles, respectively: ${\bf d}_i = {\bf r}_{i+1}
- {\bf r}_i$, $\cos \theta_i = - {\bf d}_i.{\bf d}_{i+1}/|{\bf d}_i|.|{\bf
d}_{i+1}|$. The dihedral angle $\alpha$ is defined as zero in the {\em cis}\/
conformation:
\begin{equation}
\cos \alpha = \frac{({\bf d}_1 \wedge {\bf d}_2).({\bf d}_2 \wedge {\bf d}_3)}
{|{\bf d}_1 \wedge {\bf d}_2|.|{\bf d}_2 \wedge {\bf d}_3|}.
\label{eq:alpha}
\end{equation}
All parameters are taken from Ref.~\cite{depaepe} for sake of comparison. We
recall them for clarity: $d_1^* = d_3^* = 1.54$~\AA, $d_2^*=1.52$~\AA, $k_s=
1882.8$~kJ~mol$^{-1}$~\AA$^{-2}$, $\theta_1^* = \theta_2^* = 1.937$~rad, $k_b=
376.56$~kJ~mol$^{-1}$~rad$^{-2}$, $u_3=6.6944$~kJ~mol$^{-1}$.
The LJ parameters for the $u_b(r_{14})$ term in equation (\ref{eq:but}) are
$\varepsilon=0.4184$~kJ~mol$^{-1}$ and $\sigma=3.385$~\AA. For CH$_3$, CH$_2$,
and CCl$_4$, they are respectively $\varepsilon=0.8661$, $0.4778$, and
$3.087$ kJ~mol$^{-1}$, and $\sigma=3.775$, $3.982$, and $5.27$~\AA. The
parameters for mixed interactions are obtained from the Lorentz-Berthelot
combination rules.

Monte Carlo simulations were carried out at room temperature $T=300$~K, for the
n-butane molecule in gas phase and in CCl$_4$ solvent at density
$1.614$~g~cm$^{-3}$. As in Ref.~\cite{depaepe}, 123 carbon tetrachloride
molecules were placed in a cubic box of side $L=27.124$~\AA, periodic
boundary conditions were implemented through the minimum image convention. All
LJ interactions were truncated at 11.75~\AA.

\begin{figure}[htb]
\figps{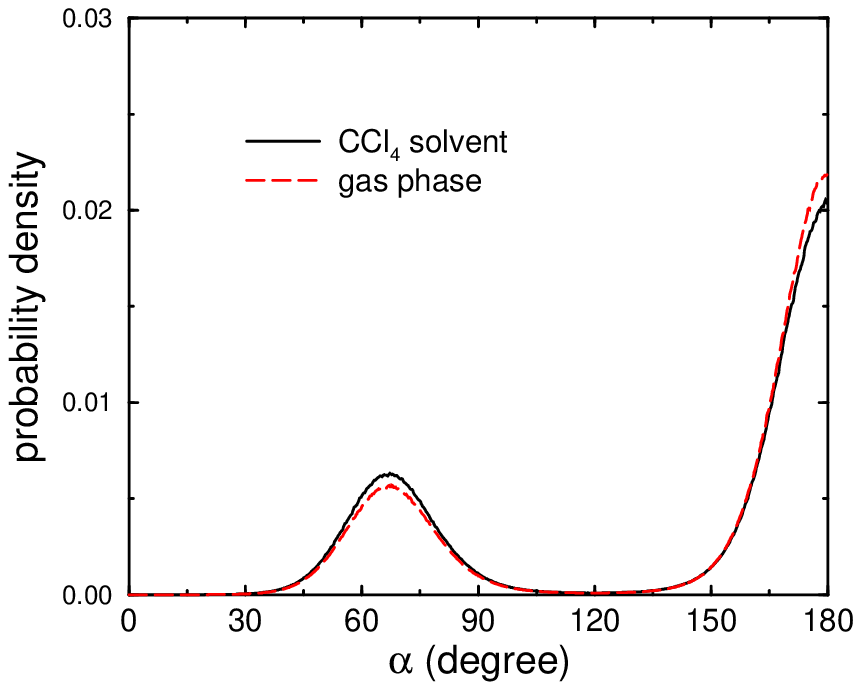}
\caption{Normalised probability density of the dihedral angle in the n-butane
molecule at $T=300$~K, in carbon tetrachloride solvent (solid line) and in gas
phase (dashed line).}
\label{fig:butane}
\end{figure}

The reaction coordinate is the dihedral angle $\alpha$, and all simulations
consisted of 20 iterations of $1.2\times 10^6$ MC cycles each, including
$2\times 10^5$ initial thermalisation cycles. 1000 bins were used for the
histogram in $\alpha$. In the case of the solvated molecule we have
supplemented the present algorithm with the preferential sampling scheme of
Owicki and Scheraga \cite{owicki} as implemented by Mehrotra and coworkers
\cite{mehrotra}. Each carbon group of the butane molecule was chosen for
displacement with probability $w_i=1/8$. The probability of choosing a solvent
molecule $i$ decays with its distance from the centre of mass of the solute
molecule $r_{0i}$ as $w_i = \gamma/r_{0i}^2$, where $\gamma$ is determined by
the normalisation of the total probabilities. Because of this bias, the
acceptance probability (\ref{eq:bdf}) is modified and becomes \cite{owicki}
\begin{equation}
\mbox{acc}({\bf R}_{\rm old}\to {\bf R}_{\rm new}) = \min \left[
1, \frac{w_{\rm new}} {w_{\rm old}} \exp(-\beta \Delta F)\right].
\label{eq:bdfps}
\end{equation}
Preferential sampling is particularly useful in this problem, because we focus
on a property of the solute molecule influenced by the solvent. The probability
densities of the dihedral angle $g(\alpha)$ in gas phase and CCl$_4$ solvent
are plotted in figure \ref{fig:butane}. The shape is very similar to the curves
reported by Depaepe {\em et al.} \cite{depaepe}, but quantitative comparison
must be made on specific indicators. We have calculated the fraction of {\em
trans}\/ isomer, $\chi_T = \int_{120^\circ}^{180^\circ} g(\alpha)d\alpha$, and
the absolute probability densities $g(\alpha^\dagger)$ at the free energy
barrier $\alpha^\dagger=120^\circ$. The results are given in table \ref{table}.
\begin{table}[htb]
\begin{center}
\begin{tabular}{lcc}
\hline
& $\chi_T$ & $g(\alpha^\dagger) \times 10^5$/deg$^{-1}$ \\
\hline
Gas phase && \\
~~Present work & $0.664\pm 0.002$ & $9.42\pm 0.07$ \\
~~Umbrella sampling$^a$ & $0.659\pm 0.004$ &$9.49\pm 0.10$
\bigskip\\
CCL$_4$ solvent && \\
~~Present work & $0.670\pm 0.011$ & $9.69\pm 0.19$ \\
~~Umbrella sampling$^a$ & $0.671\pm 0.016$ &$9.86\pm 0.36$\\
~~Blue-moon ensemble$^a$ & $0.668\pm 0.020$ &$9.68\pm 0.35$\\
\hline
$^a$ Reference \protect \cite{depaepe} &&
\end{tabular}
\end{center}
\caption{Distribution of {\em trans}\/ conformation of n-butane at equilibrium.
$\chi_T=\int_{120^\circ}^{180^\circ} g(\alpha)d\alpha$ is the fraction of
{\em trans}\/ isomer, $g(\alpha^\dagger)$ is the absolute probability
density of finding the system at the free energy barrier $\alpha^\dagger
=120^\circ$. The errors correspond to one standard deviation.}
\label{table}
\end{table}
The agreement between the work by Depaepe and coworkers and the present
results using the Wang-Landau method is quite remarkable, especially
considering that the numerical cost of the Monte Carlo simulation is lower than
the molecular dynamics simulation of Ref.~\cite{depaepe}. This quantitative
agreement suggests that the Wang-Landau method is accurate even for large
molecular systems.

\subsection{Cube-ring isomerisation in a rocksalt cluster}

\begin{figure}[htb]
\figps{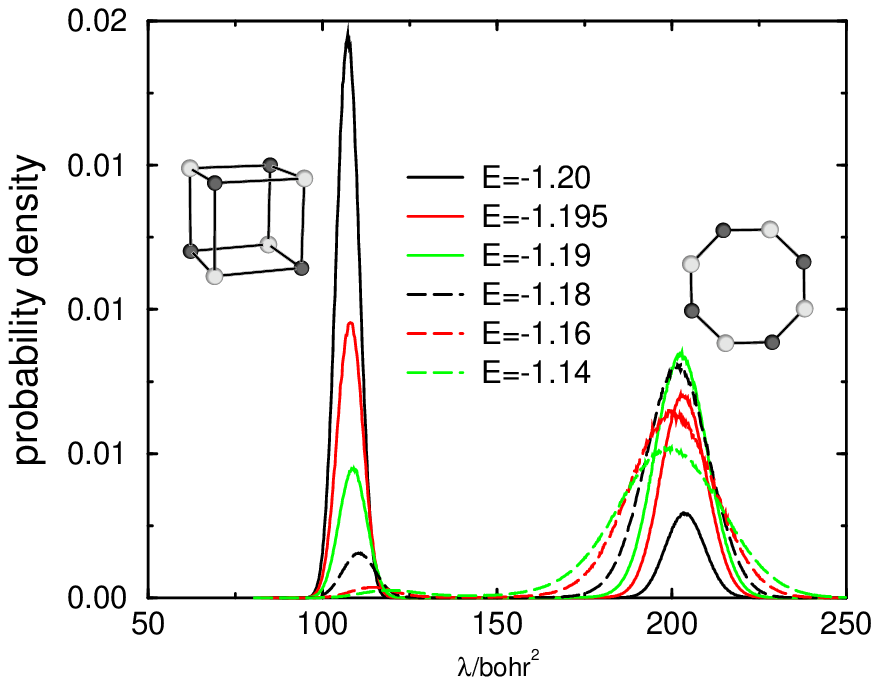}
\caption{Normalised probability density of the reaction coordinate [equation
\protect(\ref{eq:lambdanaf})] in (NaF)$_4$, in the microcanonical ensemble at
total energies in the range $-1.2\leq E\leq -1.4$ atomic units. The isomers
corresponding to each of the two peaks are also represented.}
\label{fig:naf1}
\end{figure}

\begin{figure}[htb]
\figps{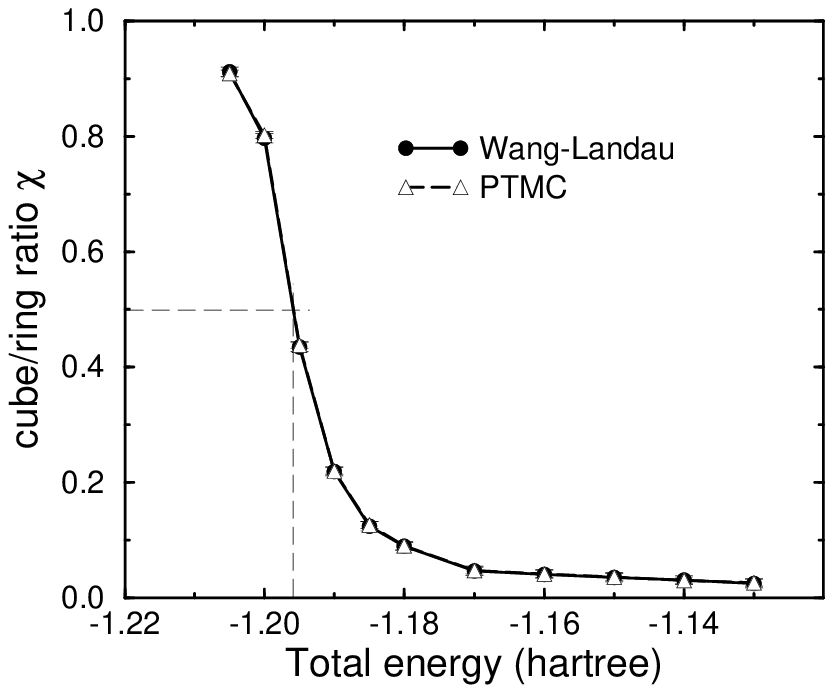}
\caption{Fraction of the cube isomer $\chi=\int_{80}^{140} g(\lambda)d\lambda$
of (NaF)$_4$ in microcanonical equilibrium. The results of the Wang-Landau
method (full circles) are compared with the results of parallel tempering
Monte Carlo (empty triangles). The dashed lines indicate the equilibrium point
where both fractions are 50\%.}
\label{fig:naf2}
\end{figure}

The previous results have shown that free energy profiles can be efficiently
calculated with the multiple range random walk technique. Data of comparable
accuracy could be obtained with the adiabatic free energy dynamics of Rosso
and coworkers \cite{rosso1,rosso2}, which has been especially designed for use
in the canonical ensemble. We now illustrate our method on another problem of
molecular character, but in the microcanonical ensemble. The small (NaF)$_4$
ionic cluster has two stable isomers, a cube ($T_d$ symmetry, energy
$E=-1.2317$~Hartree), and an octogonal ring ($C_{4v}$, $E=-1.2192$~Hartree).
The cube$\rightleftharpoons$ring isomerisation dynamics has been previously
investigated in a similar (NaCl)$_4$ cluster by Heidenreich, Oref, and Jortner
\cite{jortner}. This system is made of 8 Na$^+$ and F$^-$ ions interacting via
Coulomb forces and Born-Mayer short-range repulsion:
\begin{equation}
V({\bf R}) = \sum_{i<j} A_{ij} \exp(-\rho_{ij}r_{ij}) + \frac{q_iq_j}{r_{ij}}.
\label{eq:naf}
\end{equation}
The parameters are taken from {\em ab initio}\/ calculations \cite{abinitio}:
$A_{++} = A_{--} = A_{+-} = 41.777203$~Hartree, $\rho_{++} = \rho_{--} =
\rho_{+-} = 0.517745$~bohr$^{-1}$.
A suitable reaction coordinate is required to
distinguish the two isomers at a given total energy. After several attempts,
we have found that the principal momenta of inertia could achieve this goal.
We choose the following reaction coordinate:
\begin{equation}
\lambda({\bf R}) = \sum_i | {\bf r}_i - {\bf r}_G|^2,
\label{eq:lambdanaf}
\end{equation}
where ${\bf r}_G$ is the position of the cluster centre of mass. $\lambda$ has
roughly the value 100~bohr$^2$ in the cube isomer, and about 200~bohr$^2$ in
the ring isomer. The MC simulations were still performed for 20 iterations
of $1.2\times 10^6$ cycles each, with 1000 bins in the range $80\leq \lambda
\leq 250$ for the histograms. The probability densities of $\lambda$ in the
microcanonical ensembles at total energies $E=-1.2$, $-1.195$, $-1.19$,
$-1.18$, $-1.16$, and $-1.14$ Hartree are plotted in figure~\ref{fig:naf1}. The
ring isomer is much more ``flexible'' than the cubic isomer, which is reflected
in the relative widths of the two peaks in each of these distributions. The
fraction of cubic isomers continuously decreases in this energy range, and this
can be quantified using $\chi=\int_{80}^{140} g(\lambda)d\lambda$. In figure%
~\ref{fig:naf2} we represent the variations of $\chi$ versus total energy. For
comparison, we have also calculated $\chi$ from parallel tempering Monte Carlo
\cite{ptmc} in the microcanonical ensemble \cite{ptmcnve}. For this, we also
recorded the probability distributions of $\lambda$, and then we extracted
$\chi$. The two methods give very similar results, within the error bars, and
show that the cube and ring isomers are in equal ratios at $E\simeq -1.196$
Hartree. From this study, it could be also possible, in principle, to estimate
transition rates that could be compared with actual molecular dynamics
simulations. However, one must be careful that molecular dynamics simulations
do not strictly sample the microcanonical ensemble. Actually a geometrical
factor, which depends on the inertia momenta, hence on $\lambda$, must be
incorporated in the equilibrium statistical distribution $\rho({\bf R})$
\cite{calvo}. In order to compare with MD, the present MC simulations should be
carried out again with this extra factor in the acceptance probabilities.

\section{Conclusion}
\label{sec:ccl}

The Wang-Landau method is successfully applied to the problem of calculating
free energy profiles along reaction coordinates. This algorithm has several
advantages over most other schemes currently in use. First, it does not
require any use of constraints, which often imply complex variable
transformations and the calculation of the associated Jacobian. The method
presented in this paper can be used with any kind of reaction coordinate,
cartesian coordinates or angles. Secondly, we do not have to guess the shape
of the potential of mean force to guide the simulation with a biasing potential
as in umbrella sampling. Instead, the method has its only input in the range
of expected values of the reaction coordinate. It can subsequently provide the
guiding potential which yields a uniform probability distribution. This ``flat
histogram'' allows one to get accurate estimates of various properties using
reweighting techniques, especially when large free energy barriers are present.
Thirdly, and in contrast with the adiabatic free energy dynamics of Rosso and
coworkers \cite{rosso1,rosso2}, no unknown parameter must be estimated for
the method to work optimally. Fourthly, because it is for use with Monte Carlo
simulations, it can take benefit of many ideas aiming at accelerating
convergence, such as the preferential sampling scheme employed here for a
solvated molecule. Lastly, but importantly, the method is not limited to any
specific statistical ensemble, since it requires only the knowledge of the
equilibrium statistical distribution. The very limited input of this algorithm
makes it suitable for complex atomic and molecular systems, for which defining
a suitable reaction coordinate can already be a difficult problem.

We investigate three classes of model problems, namely a simple ten-dimensional
double-well potential energy surface, the mean force of the dihedral angle in
the gas phase and solvated n-butane molecule, and the
cube$\rightleftharpoons$ring isomerisation of the rocksalt (NaF)$_4$ cluster.
We obtain good agreement with previous works, when published data are
available. The method can be straightforwardly extended to treat free-energy
hypersurfaces by considering multidimensional weighting functions $g(\lambda_1,
\dots,\lambda_p)$ and the corresponding histograms. This could also help in
reaching equilibrium along coordinates, which are normal to the reaction
coordinate. It is also a simple matter
of algebra to generalise it to quantum free energy profiles by replacing the
conventional Boltzmann factors by Feynman path integrals in the framework of
quantum Monte Carlo.

\end{document}